\documentclass[trackchanges]{aastex701}

\usepackage{booktabs}

\newcommand{\astropyMemMapRate}{303.1}         % Astropy cutouts per second
\newcommand{\cutanaSingleRate}{878.3}          % Cutana single thread cutouts/sec
\newcommand{\cutanaFourRate}{2264.0}           % Cutana 4 threads cutouts/sec
\newcommand{\speedupSingle}{2.90}              % Single thread speedup factor (878.3/303.1)
\newcommand{\speedupFour}{7.47}                % Four thread speedup factor (2264.0/303.1)
\newcommand{\scalingFactor}{2.58}              % Cutana single vs 4-thread scaling (2264.0/878.3)
\newcommand{\scalingEfficiency}{64.5}          % Cutana scaling efficiency 1->4 threads (%)
\newcommand{\memoryUsageSingle}{2.49}          % Memory usage single thread peak (GB)
\newcommand{\memoryUsageFour}{7.59}            % Memory usage 4 threads peak (GB)
\newcommand{\memoryUsagePeak}{2.49}            % Peak memory footprint of loading a tile (GB)
\newcommand{\memoryUsageAverage}{1.8}          % Average memory footprint of loading a tile (GB)
\newcommand{\sparceCutoutsSecond}{60.5}        % Cutouts/second in sparse tile (32t, 4 FITS)
\newcommand{\denseCutoutsSecond}{878.3}        % Cutouts/second in dense tile (8t, 1 FITS)
\newcommand{\astropyMemoryPeak}{9.26}          % Astropy 4 threads peak memory (GB)

\newcommand{\cutana}{\texttt{Cutana}}      % Commands for format of software names in AAS Journals.
\newcommand{\euclid}{\textit{Euclid}}
\newcommand{\astropy}{\texttt{Astropy}}

\newcommand{\todo}[1]{\iffalse #1 \fi}

\todo{Add mroe and better references @David}
\todo{Add actual results @Pablo}

\begin{document}

\title{Cutana: A High-Performance Tool for Astronomical Image Cutout Generation at Petabyte Scale}

\author[0000-0002-5631-8240]{Pablo G\'{o}mez}
\affiliation{European Space Agency (ESA), European Space Astronomy Centre (ESAC), Camino Bajo del Castillo s/n, 28962, Villanueva de la Ca\~{n}ada, Madrid, Spain}
\email{pablo.gomez@esa.int}

\author[0009-0003-3810-1245]{Laslo Erik Ruhberg}
\affiliation{Astronomisches Rechen-Institut (ARI), Zentrum f\"{u}r Astronomie, Universit\"{a}t Heidelberg, M\"{o}nchhofstr. 12-14, 69120 Heidelberg, Germany}
\email{laslo.ruhberg@uni-heidelberg.de}

\author{Kristin Anett Remmelgas}
\affiliation{European Space Agency (ESA), European Space Astronomy Centre (ESAC), Camino Bajo del Castillo s/n, 28962, Villanueva de la Ca\~{n}ada, Madrid, Spain}
\email{kristin.remmelgas@esa.int}

\author[0000-0003-1217-4617]{David O'Ryan}
\affiliation{European Space Agency (ESA), European Space Astronomy Centre (ESAC), Camino Bajo del Castillo s/n, 28962, Villanueva de la Ca\~{n}ada, Madrid, Spain}
\email{david.oryan@esa.int}

\begin{abstract}
The \euclid{} Quick Data Release 1 (Q1) encompasses 30 million sources across 63.1 square degrees, marking the beginning of petabyte-scale data delivery through Data Release 1 (DR1) and subsequent releases. Systematic exploitation of such datasets requires extracting millions of source-specific cutouts, yet standard tools like \astropy{}'s Cutout2D process sources individually, creating bottlenecks for large catalogues. We introduce \cutana{}, a memory-efficient software tool optimised for batch processing in both local and cloud-native environments. \cutana{} employs vectorised \texttt{NumPy} operations to extract cutout batches simultaneously from FITS tiles, implements automated memory-aware scheduling, and supports both Zarr and FITS output formats with multiple common normalisation schemes (asinh, log, zscale). \cutana{} outperforms \astropy{} in all tested Q1 subset scenarios achieving near linear scaling and processing thousands of cutouts per second. On just four worker threads, Cutana can process all of Q1 in under four hours. The tool includes an \texttt{ipywidget} interface for parameter configuration and real-time monitoring. Integration with ESA Datalabs is underway for the \euclid{} DR1 release, with open-source release\footnote{\url{https://github.com/ESA/Cutana}} pending ESA open-source licensing processes.
\end{abstract}

\todo{Double-check these match}

\section{Introduction}
\noindent Modern astronomical surveys are entering an unprecedented era of data generation. Telescopes like ESA's \euclid{} mission \citep{2025A&A...697A...1E} and the Vera C. Rubin Observatory \citep{2019ApJ...873..111I} will produce petabyte-scale datasets of billions of sources. \euclid{}'s first Quick Data Release (Q1) already provides 63.1 square degrees of observations containing approximately 30 million objects \citep{euclidcollaboration2025euclidquickdatarelease}, foreshadowing the massive data volumes expected from the full survey. These vast archives harbour rare and scientifically valuable phenomena, as demonstrated by recent large-scale searches such as the identification of astrophysical anomalies in 99.6 million cutouts from the \textit{Hubble} Legacy Archive using the \texttt{AnomalyMatch} method \citep{2025arXiv250503508O}.

Maximising scientific return from these datasets fundamentally relies on generating source-specific image cutouts for detailed analysis. However, the computational infrastructure required to process such volumes presents significant challenges. Modern computational platforms like ESA Datalabs provide Kubernetes clusters and computational grids optimised for large-scale processing, yet these environments require following high-performance computing practices such as careful memory management, and efficient compute utilisation, which traditional astronomical tools were not specifically designed to address.

The community's standard approach using \astropy{} processes sources individually through its Cutout2D implementation, creating substantial overhead when handling millions of objects. While \astropy{} provides robust world coordinate system handling and FITS manipulation, its single-source processing paradigm can introduce bottlenecks. This leaves topics such as parallelisation and memory management to the user. This typically involves custom scripts wrapping \astropy{} with parallelisation, yet these often encounter memory management challenges in constrained environments, particularly when processing large FITS files with multiple workers.

Hence, we present \cutana{}, a purpose-built tool designed to address these limitations through vectorised, load-balanced batch processing while maintaining compatibility with astronomical data standards and formats. The software combines efficient algorithms with dynamic resource management, making it suitable for deployment in both traditional computing environments and modern containerised infrastructure. A streamlined graphical user interface enables rapid cutout creation.

\begin{figure*}[ht!]
\includegraphics[width=0.60\linewidth]{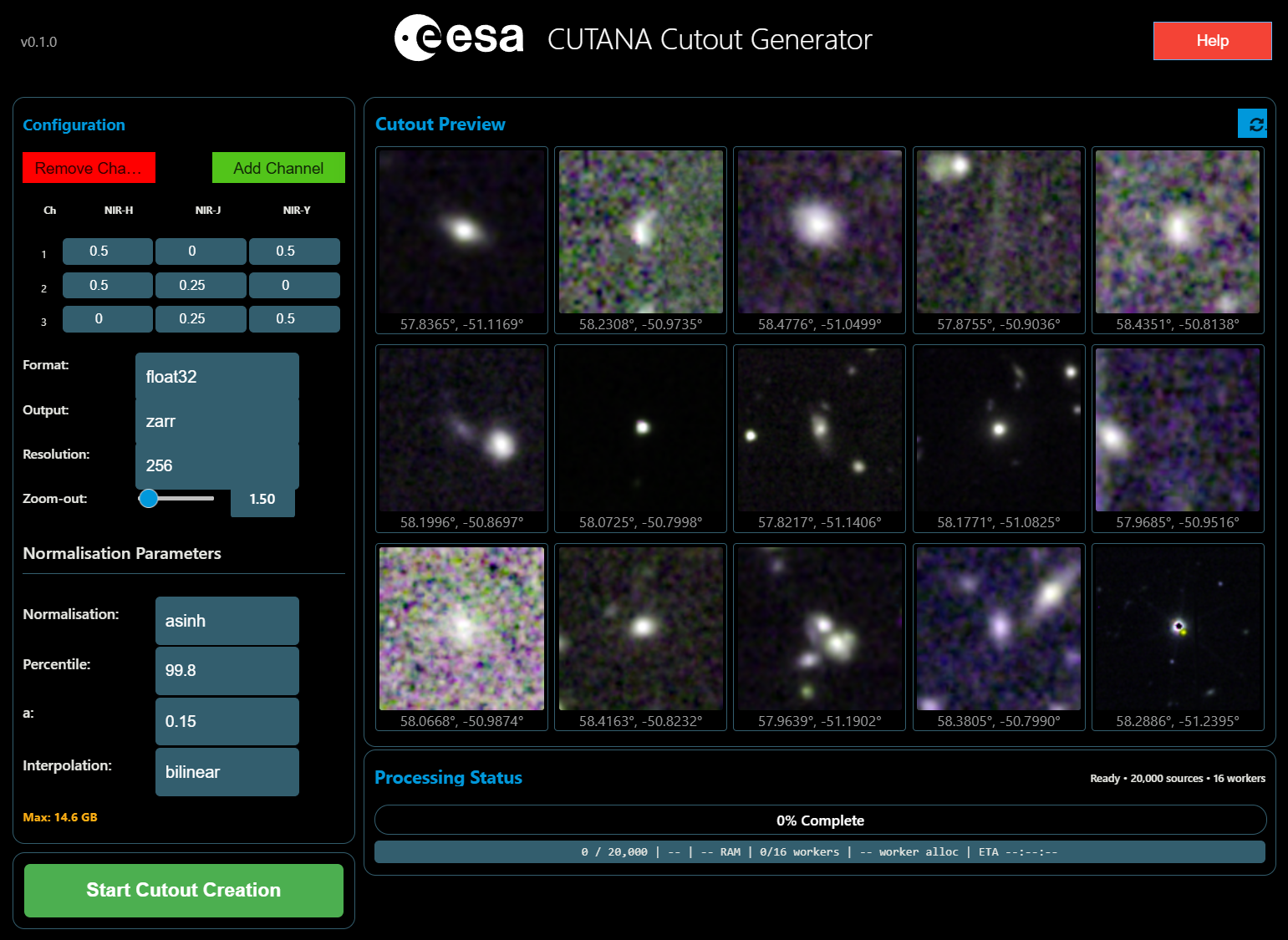}\centering
\caption{\cutana{}'s ipywidget-based interface within ESA Datalabs showing configuration panel (left) for source catalogue, output format, normalisation, and processing parameters, alongside real-time cutout preview grid (right) enabling validation of settings before large-scale processing on \euclid{} Q1 data.}
\label{fig:cutana_ui}
\end{figure*}

\section{Implementation}
\noindent \cutana{} is implemented as a modular Python package with distinct components for data processing, I/O operations, and user interaction. The architecture separates the core batch processing engine from the graphical interface, enabling both interactive usage through Jupyter notebooks and automated execution in pipeline environments.

The backend employs a multi-stage pipeline optimised for astronomical image processing. The input source catalogues provide celestial coordinates (right ascension, declination), object sizes, and FITS file paths. Rather than processing sources individually, \cutana{} groups sources by their parent tiles, enabling efficient batch extraction through vectorised \texttt{NumPy} operations. The pipeline stages include: (1) catalogue validation and source grouping by tile, (2) tile-based batch processing with memory-mapped FITS access, (3) coordinate transformation and cutout extraction using optimised array slicing, (4) normalisation and format conversion, and (5) metadata generation and output packaging.

The system implements dynamic load balancing through continuous memory monitoring, adjusting worker processes based on available resources with a configurable safety margin (default 15\%). This approach prevents memory exhaustion while maximising throughput, particularly critical in containerised environments with strict resource limits.

% \subsection{Data Formats and Storage}
\noindent \cutana{} supports both traditional FITS and the cloud-optimised Zarr format \citep{2022zndo...6637626M}. Zarr provides efficient storage for multidimensional arrays through chunked and compressed storage that enables parallel I/O operations, making it ideal for cloud computing environments. Output data includes comprehensive metadata in parquet format. The metadata preserves source identifiers, coordinates, processing timestamps, and array indices, enabling rapid source retrieval from batch-processed archives.

The tool leverages two additional open-source packages developed by our team: \texttt{fitsbolt}\footnote{\url{https://github.com/Lasloruhberg/fitsbolt/}} for high-performance FITS manipulation and normalisation operations, and \texttt{images-to-zarr}\footnote{\url{https://github.com/gomezzz/images\_to\_zarr/}} for efficient conversion between image formats and Zarr archives. These tools provide optimised implementations of critical operations, contributing to \cutana{}'s overall performance advantages.

% \subsection{Image Processing Capabilities}
\noindent The software incorporates multiple image normalisation methods essential for astronomical analysis, powered by the \texttt{fitsbolt} library. Available stretches include asinh, log, and zscale transformations, each with configurable parameters unified under a consistent interface. The system supports channel blending for multi-band observations, enabling users to combine multiple FITS extensions with custom weights for composite visualisation or analysis. These normalisation options can be previewed in real-time through the user interface before processing.

Cutout extraction implements a configurable padding factor controlling the extraction area relative to source size. With that factor, both zoom-ins ($ < 1$), e.g. for crowded fields, as well as padding ($ > 1$)  , i.e. larger cutout areas, can be achieved. All cutouts undergo resampling to a user-specified target resolution (default 256×256 pixels) using configurable interpolation methods including bilinear, nearest-neighbour, cubic, and Lanczos resampling to enable efficient storage as large tensors.

\section{Results}
\noindent We evaluated \cutana{}'s performance using \euclid{} Q1 data, generating 256×256 pixel cutouts from the MER catalogue. Testing was conducted on an Intel i5-13400F system with 64 GB DDR4 RAM to ensure reproducible comparisons with \astropy{}'s Cutout2D implementation. As \astropy{} lacks native parallelisation support, we implemented single-threaded comparisons to establish baseline performance, utilising the automated parallelisation of underlying libraries via thread pinning to either one or four cores. 
% \subsection{Processing Speed and Parallel Scaling}
\noindent Table~\ref{tab:framework_comparison} presents comprehensive performance measurements across two distinct test configurations varying in tile count, FITS file complexity, and source catalogue size. The primary comparison (8 tiles, 1 FITS file, 200,000 sources) demonstrates that single-threaded \cutana{} execution achieves a \speedupSingle{}$\times$ improvement over \astropy{}'s memory-mapped implementation, processing \cutanaSingleRate{} cutouts per second compared to \astropyMemMapRate{} for \astropy{}. With four workers, \cutana{} reaches \cutanaFourRate{} cutouts per second, representing a \speedupFour{}$\times$ speed-up over the baseline single-threaded \astropy{} implementation.

\begin{deluxetable*}{cccc}
\tabletypesize{\small}
\tablewidth{0.95\textwidth}
\tablecaption{Framework Comparison: \astropy{} vs \texttt{Cutana} Performance \label{tab:framework_comparison}}
\tablehead{
\colhead{Tiles - Channels (i.e. FITS files) - Total \# of sources} &
\colhead{Tool (Threads / Workers)} &
\colhead{Runtime (s)} &
\colhead{Sources/sec}
} 

\renewcommand{\arraystretch}{0.9} % <-- reduces row spacing (default is 1.0)
\setlength{\tabcolsep}{3pt}       % <-- reduces horizontal padding in columns

% \colnumbers
\startdata
8 Tiles - 1 Channel - 200,000 & \astropy{} (1 Thread) & 659.9 & 303.1 \\
8 Tiles - 1 Channel - 200,000 & \texttt{Cutana} (1 Worker) & 227.7 & 878.3 \\
8 Tiles - 1 Channel - 200,000 & \astropy{} (4 Threads) & 626.1 & 319.4 \\
8 Tiles - 1 Channel - 200,000 & \texttt{Cutana} (4 Workers) & 88.3 & 2264.0 \\
\hline
32 Tiles - 4 Channel - 32,000 & \astropy{} (1 Thread) & 208.0 & 153.9 \\
32 Tiles - 4 Channel - 32,000 & \texttt{Cutana} (1 Worker) & 529.2 & 60.5 \\
32 Tiles - 4 Channel - 32,000 & \astropy{} (4 Threads) & 98.3 & 325.5 \\
32 Tiles - 4 Channel - 32,000 & \texttt{Cutana} (4 Workers) & 69.6 & 460.0
\enddata
\end{deluxetable*}

The performance gains derive from three primary optimisations: (1) vectorised batch processing eliminating per-source overhead, (2) efficient memory management through tile-based grouping, and (3) optimised I/O patterns minimising disk access latency. The batch processing approach proves particularly effective for dense source catalogues where many cutouts originate from the same tile, as evidenced by the 8-tile configuration achieving the highest throughput rates.
\cutana{} demonstrates scaling from single to multi-threaded execution, with a scaling factor of \scalingFactor{}$\times$ when moving from one to four workers for the 8 Tile Scenario. This represents a parallel efficiency of \scalingEfficiency{}\%. The sub-linear scaling primarily results from I/O contention when multiple workers access the same FITS files simultaneously. This is noticeable as thread scaling efficiency is super-linear (7.6x sources/sec with workers) for the 32 Tiles Scenario.
The two test configurations reveal how number of sources per tile affects processing throughput. The dense configuration (8 tiles, 1 FITS file, 200,000 sources) achieves \denseCutoutsSecond{} cutouts/second with single-worker \cutana{}, demonstrating optimal performance when many sources originate from few tiles. In contrast, the sparse configuration (32 tiles, 4 FITS files, 32,000 sources) demonstrates \sparceCutoutsSecond{} cutouts/second, a reduction attributable to increased tile-loading overhead. This behaviour reveals that \cutana{} operates most efficiently when compute-bound rather than I/O-bound, whilst \astropy{} exhibits limited performance variation across these scenarios.

\begin{figure*}[ht!]
\includegraphics[width=0.9\linewidth]{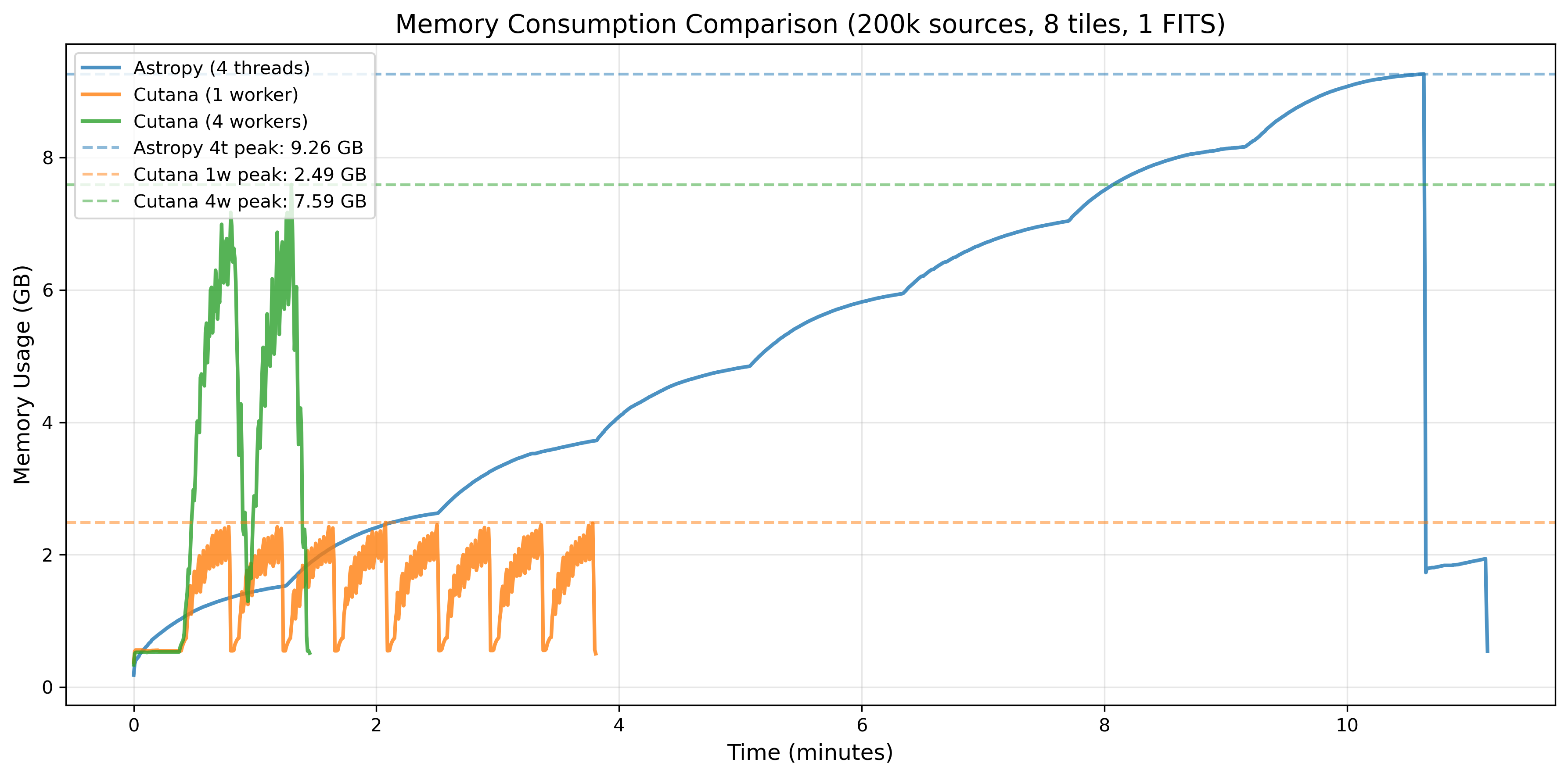}\centering
\caption{Memory consumption comparison processing 200,000 sources across 8 tiles. \astropy{} (blue) exhibits monotonic growth, whilst \cutana{} (orange: 1 worker, green: 4 workers) maintains bounded memory through tile-wise recycling. The sawtooth pattern reflects controlled loading cycles preventing exhaustion whilst maintaining throughput.}
\label{fig:memory}
\end{figure*}

% \subsection{Memory Management and Resource Efficiency}
\noindent Figure~\ref{fig:memory} illustrates the critical distinction between \cutana{}'s dynamic memory management and \astropy{}'s linear accumulation pattern. When processing 200,000 sources with naive parallelisation, \astropy{} exhibits continuous memory growth, reaching \astropyMemoryPeak{} GB before completion. This behaviour stems from per-source processing accumulating intermediate arrays and Python objects without garbage collection opportunities between cutouts.

In contrast, \cutana{}'s tile-based architecture with worker process recycling maintains bounded memory consumption regardless of catalogue size. Single-worker execution peaks at \memoryUsageSingle{} GB, whilst four-worker parallel processing reaches \memoryUsageFour{} GB. The characteristic sawtooth pattern in \cutana{}'s memory profile reflects deliberate batch processing cycles: each worker receives a job for ideally one tile (minimum and maximum numbers of sources can be specified to minimise overhead), processes all associated sources in batches, writes outputs, and terminates before a new one worker is spawned for the next job. This memory-efficient design proves essential for containerised deployments where resource limits are strictly enforced. 

Peak memory usage of \memoryUsagePeak{} GB per worker during tile loading phases, with average footprint of \memoryUsageAverage{} GB during steady-state processing, enables predictable resource allocation. This controlled memory behaviour, combined with automatic management of number of active workers, prevents out-of-memory failures that commonly plague large-scale processing jobs whilst maintaining near-optimal throughput through intelligent load balancing.

\section{Conclusion}
\cutana{} addresses the critical need for scalable, memory-efficient cutout generation in the era of petabyte-scale astronomical surveys. By implementing batch processing, dynamic load balancing, and support for efficient data formats, the software tool enables efficient processing of massive datasets in state-of-the-art computational environments. The demonstrated performance improvements combined with robust memory management make \cutana{} suitable for integration into survey data processing pipelines, as evidenced by its deployment within ESA Datalabs for \euclid{} data releases.

As astronomical data volumes continue to grow, tools like \cutana{} become essential infrastructure for scientific discovery. By bridging the gap between traditional astronomical software and modern computational paradigms, we enable the community to fully exploit the scientific potential of current and future surveys, ultimately advancing our understanding of the cosmos through efficient data exploration and analysis.

\begin{acknowledgments}
We thank the \euclid{} Consortium for providing access to Q1 data and valuable feedback during development. We acknowledge ESA Datalabs for computational resources and platform support. Software development and manuscript preparation made use of AI tools.
\end{acknowledgments}

\bibliography{references}
\bibliographystyle{aasjournalv7}

\end{document}